\documentclass{article}

\usepackage[final]{neurips_2023}

\usepackage[utf8]{inputenc} 
\usepackage[T1]{fontenc}    
\usepackage{hyperref}       
\usepackage{url}            
\usepackage{booktabs}       
\usepackage{amsfonts}       
\usepackage{nicefrac}       
\usepackage{microtype}      
\usepackage{xcolor}         
\usepackage{graphicx}
\usepackage{listings}
\usepackage{makecell}
\usepackage{array}
\usepackage{textcomp}
\usepackage{lipsum}
\usepackage{geometry}
\usepackage{adjustbox}
\usepackage{chngpage}
\usepackage{float}
\usepackage{amsmath}

\usepackage{tikz}
\usetikzlibrary{arrows.meta}
\usetikzlibrary{positioning}
\tikzset{%
  >={Latex[width=2mm,length=2mm]},
            base/.style = {rectangle, rounded corners, draw=black,
                           minimum width=2cm, minimum height=0.5cm,
                           text centered, font=\sffamily, text width=3cm, font=\small},
  static/.style = {base, fill=yellow!30},
       chatgpt/.style = {base, fill=red!10},
    docker/.style = {base, fill=red!30},
    done/.style = {base, fill=green!30},
}

\setcitestyle{square,numbers}

\title{Proof-of-concept: Using ChatGPT to Translate and Modernize an Earth System Model from Fortran to Python/JAX}

\author{
  Anthony Zhou \\
  Columbia University\\
  New York, NY 10027 \\
  \texttt{az2681@columbia.edu} \\
  \And
  Linnia Hawkins \\
  Columbia University\\
  New York, NY 10027 \\
  \texttt{lh3194@columbia.edu} \\
  \And
  Pierre Gentine \\
  Columbia University \\
  New York, NY 10027 \\
  \texttt{pg2328@columbia.edu} \\
}

\begin{document}

\maketitle

\begin{abstract}
Earth system models (ESMs) are vital for understanding past, present, and future climate, but they suffer from legacy technical infrastructure. ESMs are primarily implemented in Fortran, a language that poses a high barrier of entry for early career scientists and lacks a GPU runtime, which has become essential for continued advancement as GPU power increases and CPU scaling slows. Fortran also lacks differentiability — the capacity to differentiate through numerical code — which enables hybrid models that integrate machine learning methods. Converting an ESM from Fortran to Python/JAX could resolve these issues. This work presents a semi-automated method for translating individual model components from Fortran to Python/JAX using a large language model (GPT-4). By translating the photosynthesis model from the Community Earth System Model (CESM), we demonstrate that the Python/JAX version results in up to 100x faster runtimes using GPU parallelization, and enables parameter estimation via automatic differentiation. The Python code is also easy to read and run and could be used by instructors in the classroom. This work illustrates a path towards the ultimate goal of making climate models fast, inclusive, and differentiable.
\end{abstract}

\section{Introduction}

Over the last decades, climate models have evolved from very coarse ocean-atmosphere models to complex Earth system models including all Earth components — the ocean, atmosphere, cryosphere and land and detailed biogeochemical cycle — \citep{edwards2011history} and are now being used to address a vast array of questions from projecting natural sinks \citep{friedlingstein2014uncertainties} to the impact of land use change on climate \citep{findell2017impact}. Over decades of model development, Earth system models have become huge computing programs written in Fortran or C, for legacy and compatibility reasons. Yet most early career scientists do not know those languages, which hinders model development. 

Further, because of their complexity, running these models is resource-intensive \citep{schar2020kilometer,palmer2014climate}. Most (perhaps all) models rely on parallelized CPU hardware, even though highly-parallelized code can now be run more efficiently on Graphics Processing Units (GPUs) \citep{baji2018evolution} or Tensor Processing Units (TPUs) \citep{cass2019taking}. Earth system modeling would dramatically benefit from hardware acceleration, so that models could be run at higher resolutions and resolve important fine-scale processes like ocean eddies or deep convection. Using high-level languages also makes model code robust to future hardware evolution; this type of (r)evolution is already starting in other fields of physics \citep{campagne2023jax,bezgin2023jax}. 

Finally, Earth System Models are starting to use machine learning (ML) to represent subgrid processes \citep{rasp2018deep,gentine2018could,yuval2020stable,bolton2019applications}. Because of their native Python infrastructure, those machine learning components are not straightforward to implement in Fortran- or C/C++-based Earth System Models and require a bridge (e.g., Fortran to Python), without the flexibility to do online learning (updating between ML and the physical code while the physical model runs) \citep{ott2020fortran} \citep{saul2004overview}. The missing piece is automatic differentiation, which allows us to take the partial derivatives to any parameters in the code. Python has many libraries supporting differentiation; one of these is JAX, which enables automatic differentiation and hardware acceleration of Python code while maintaining a near-Python visual look \cite{jax2018github}.

In this work, we intend to break this Earth System Model language and hardware deadlock by modernizing Earth System Model codes to run on modern hardware (e.g., GPUs) and enabling automatic differentiation \citep{margossian2019review} using Python and JAX to enhance both efficiency and precision \citep{jax2018github}, while also making model development more inclusive. To realize this goal, our approach leverages recent developments in large-language models (LLMs) to rapidly translate code from one language to another \citep{kasneci2023chatgpt} \citep{weisz2022better}. Specifically, we split the model into units using static analysis, develop a topological sort order based on dependencies, and translate each unit using OpenAI's GPT-4 \citep{sanderson2023gpt}. 

Our contribution is twofold: we demonstrate the Python/JAX version's utility through an example model (leaf-level photosynthesis in the Community Land Model (CLM) of CESM), and we offer open source tools to address translation challenges. As an added benefit, we also demonstrate that the Python translation can be computationally more efficient than the original model by leveraging GPUs with the JAX library. 

\section{Translation Workflow}
\label{translation}


Large language models alone cannot translate a whole codebase for two reasons. First, there are more tokens (i.e., words) in a typical codebase than a model like GPT-4 can accept in its token limit. Second, the model alone often writes incorrect code. To resolve these issues, we've implemented a divide-and-conquer approach that splits a Fortran codebase into ordered units using static analysis, generates Python/JAX translations using the GPT-4 API, and uses unit test outputs to iterate on the generated code. Figure \ref{fig:workflow} shows the architecture of our translation process (see details in appendix). 

\begin{figure}[ht]
  \centering
  \begin{tikzpicture}[node distance=0.5cm,
    every node/.style={fill=white, font=\sffamily}, align=center]
  \node (start)             [static]              {Isolate Fortran source code and dependencies};
  \node (unitTestsBlock)    [chatgpt, below=of start]                   {Generate Fortran unit tests};
  \node (pythonTestsBlock)     [chatgpt, right=of unitTestsBlock]   {Generate Python unit tests};
  \node (updateTestsBlock)      [chatgpt, right=of pythonTestsBlock]
                                                      {Use unit test output to update unit tests and code};
  \node (generatePythonBlock)      [chatgpt, below of=pythonTestsBlock, yshift=-1cm]
                                                                {Generate Python code};

\node (testsPass)      [docker, below of=updateTestsBlock, yshift=-1cm]
                                                                {Do unit tests pass?};                                                              
\node (done)      [done, below of=testsPass, yshift=-1cm]
                                                                {Done};                                                              
  \draw[->]             (start) -- (unitTestsBlock);
  \draw[->]     (unitTestsBlock) -- (pythonTestsBlock);
  \draw[->]      (pythonTestsBlock) -- (generatePythonBlock);
  \draw[->]     (updateTestsBlock) -- (pythonTestsBlock);
  \draw[->]     (generatePythonBlock) -- (testsPass);
  \draw[->]     (testsPass) -- (updateTestsBlock);
  \draw[->]     (testsPass) -- (done);
  \draw[->]     (done) -- ++(-10,0) -- ++(0, 1) |- (start.west);
\end{tikzpicture}
  \caption{Workflow for translating a climate model from Fortran to Python, using static analysis, code generation, and unit testing. }
  \label{fig:workflow}
\end{figure}
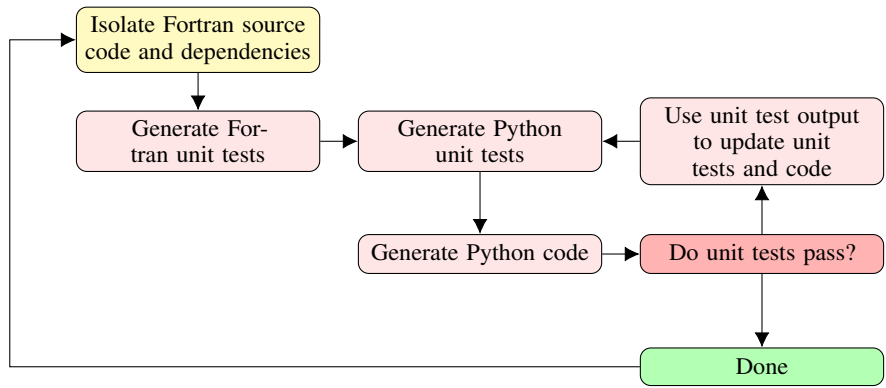

\section{Translation Evaluation}
\label{evaluation}

\subsection{Runtime}

\begin{figure}[ht]
  \centering
  \includegraphics[width=0.8\textwidth]{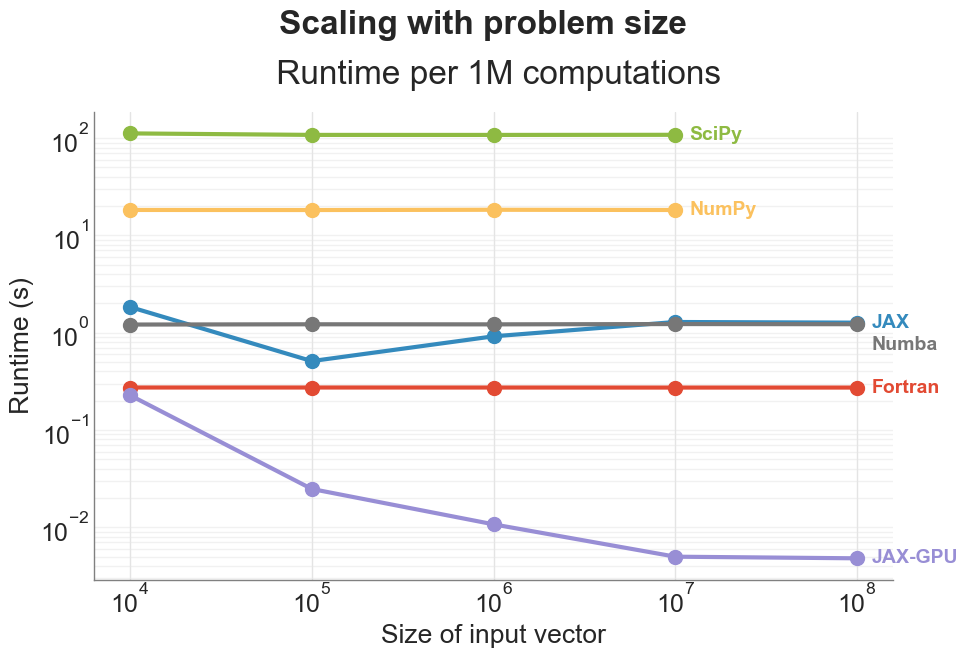}
  \caption{Comparing runtime of leaf-level photosynthesis in several Python translations with the original Fortran version. Runtime was measured on an Amazon EC2 G5.4xlarge instance with one NVIDIA A10G GPU. }
  \label{fig:ci_runtime}
\end{figure}

We made several versions of the original photosynthesis code, which were mostly LLM-generated. There are four Python versions: \verb|NumPy| makes a direct translation of the algorithm from Fortran into NumPy. \verb|Numba| takes \verb|NumPy| and adds just-in-time (jit) compiling using Numba \citep{lam2015numba}. \verb|SciPy| uses a SciPy library function for rootfinding \citep{virtanen2020scipy}, and \verb|JAX| uses the JAX library for jit compilation. Note that this JAX version, in contrast to the other Python versions, required substantial modification from the LLM-generated code. See Figure \ref{fig:ci_runtime} for a visual comparison. 

The code we translated takes as input an initial value of internal partial pressure of CO2 in the leaf, and iteratively solves for the x-intercept of a function for stomatal conductance based on this guess \citep{medlyn2011reconciling} \cite{CLM50_Tech_Note}. To generate these runtime results, we ran each of the versions (4 Python versions + 1 on GPU, 1 Fortran version) using vectors sampling from a range of input values from 35 to 70 Pa. 

From the runtime results (smaller is better), we observe that \verb|JAX-GPU| was the fastest, with \verb|Fortran| a full two orders of magnitude slower. \verb|JAX| on CPU and \verb|Numba| (both jit-compiled) are slightly slower than Fortran. \verb|NumPy| and \verb|SciPy| are both even slower. From this we can conclude two things. First, these results show that GPU parallelization on a GPU can lead to significant runtime improvements relative to sequential Fortran on a roughly equivalent CPU (included with the same machine). Second, even \verb|JAX-CPU| and \verb|Numba| perform reasonably well compared to Fortran, suggesting that jit-compilation alone can make up for much of the speed difference between Fortran and Python. 

\subsection{Parameter Estimation}

While the equations describing photosynthesis are well understood, there are many uncertain parameters that vary by plant species or even leaf position in the canopy. To demonstrate the usefulness of automatic differentiation in Python, we chose one parameter to estimate: the maximum rate of carboxylation, Vcmax, which plays a key role in determining the co-limited rate of assimilation. 

One method to obtain the optimal set of parameters, is by running a strategically sampled parameter perturbation experiment, where multiple parameters are varied simultaneously to identify the best set of parameters (e.g., \cite{williamson2015}; \cite{hourdin2017}; \cite{couvreux2021}). However, studies have shown that it is challenging to find global optimal parameter settings that improve overall skill in a climate model (e.g., \cite{mcneall2016}; \cite{dagon2020}). An alternative method, inspired by machine learning, is using gradient descent to find parameter values that align the model with measured data \citep{ruder2016overview}. Gradient descent has become an increasingly optimized operation in Python, thanks to techniques like minibatches and stochastic methods \citep{bottou2012stochastic}. 

\begin{figure}[ht]
  \centering
  \includegraphics[width=0.8\textwidth]{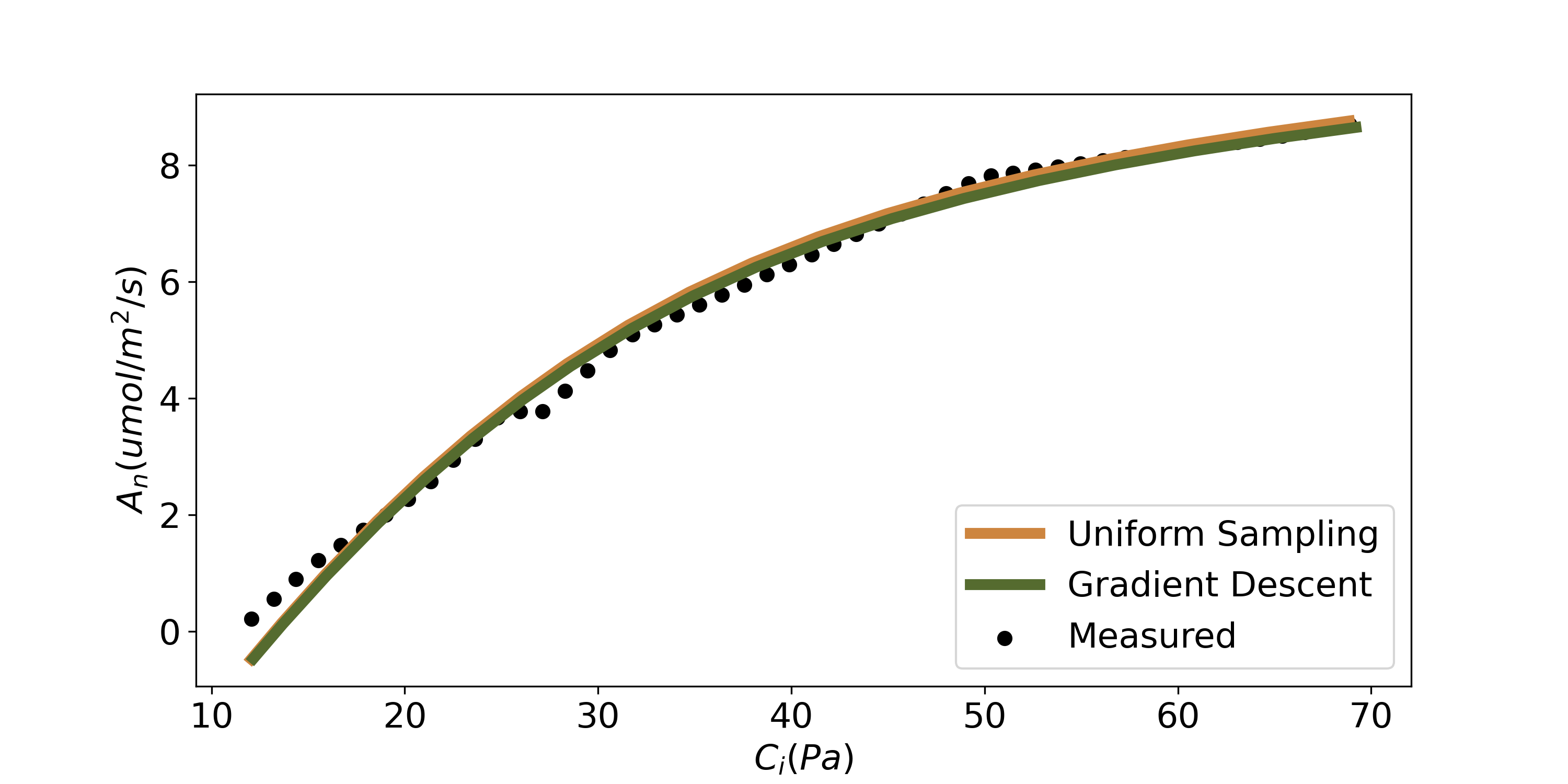}
  \caption{Measured (points) and modeled (lines) relationship between the internal partial pressure of CO2 (Pa) and the rate of assimilation (umol/m2/s). The modeled values use the Vcmax parameter value selected using either uniform sampling (orange) or gradient descent (green). }
  \label{fig:param_estimation}
\end{figure}

In Figure \ref{fig:param_estimation}, we compared parameter estimation methods for optimizing the Vcmax parameter using observed measurements of assimilation rate (An) at a range of partial pressures of CO2 within leaf (observations from a Ponderosa pine tree at the US-Me2 AmeriFlux site in Central Oregon). We employed a uniform sampling parameter perturbation scheme and gradient descent to identify the Vcmax setting that minimized the error between the model simulations and observations. In this simplified leaf-level model, uniform sampling and gradient descent converged to similar values (Vcmax=38.776 $\mu$mol $\text{CO}_{\text{2}}\text{ m}^{-2}\text{s}^{-1}$ and Vcmax=38.383 $\mu$mol $\text{CO}_{\text{2}}\text{ m}^{-2}\text{s}^{-1}$ respectively), though gradient descent took fewer iterations (10 gradient descent steps as opposed to 50 sample points in uniform sampling) and achieves a lower loss value (7.26 and 6.39 respectively in mean squared error). 

These results support the idea that gradient descent and automatic differentiation would also be more efficient when estimating more parameters for multiple modules of CLM. Our code conversion to JAX unleashes the use of ML-methods, such as stochastic gradient descent, for efficient parameter tuning through automatic differentiation \citep{baydin2018automatic}. 

\subsection{Future Work}

While implementing automatic translation, we faced challenges including poor Fortran generation quality, inaccurate unit tests, incorrect use of imported modules, and GPT-4 token limits. For scaling up translation, next steps could include using data flow and compiler representations for translation (as in \cite{szafraniec2022code} and \cite{GraphCodeBERT}), building logging to track function calls, and designing copilot-style interfaces to support human translators. Solving these engineering challenges will further decrease the amount of manual work involved in translating and modernizing climate models. 

\section{Conclusions}

Migrating a full climate model from Fortran to Python, even with language models like ChatGPT, is challenging due to important dependencies and logical errors in code generation. However, translating individual components (like leaf-level photosynthesis) from Fortran to Python/JAX is both useful and now within reach, and tools like the static analysis and language models presented in this work will make this easier. Even with just a single component translated and modernized in Python, one could experiment with parameter estimation, measuring sensitivity to parameters (quantifying parametric uncertainty and getting faster feedback loops for model development), running on GPU, and translating model components for offline experiments. 

Eventually, this work aims to pave the way for the development of a future where climate models are differentiable and GPU/TPU-friendly, making them faster and more accurate, while also written in high-level languages so that they are more inclusive of junior scientists to accelerate progress on the critical program of climate change modeling and adaptation. 

{
\small
\bibliographystyle{plainnat}
\bibliography{refs}
}

\section{Appendix}
\label{appendix}

This appendix provides details on the implementation of the translation method, useful for those who seek to build on this work. Our semi-automated translation process works in two steps: first we divide the Fortran codebase into manageable chunks using static analysis. Second, we "conquer" (translate) each of these chunks from Fortran to Python by running unit tests with iterative language model generations. 

\subsection{Prompting}

With large language models (LLMs), changing the prompt can significantly affect the generated response \citep{DBLP:journals/corr/abs-2107-13586}. Therefore, we developed a variety of prompts for the tasks of translating Fortran to Python, writing Fortran unit tests, and writing Python unit tests. The GPT-4 Chat Completion API accepts prompts in the form of chat messages \citep{gpt4apidocs}. In this work, we used the System message and one User message for each API call, as shown in Table \ref{tab:prompts}. Some sample outputs are shown in Table \ref{tab:outputs}.

\begin{table}
  \centering
  \begin{tabular}{
    >{\raggedright}p{2.5cm}
    >{\raggedright}p{5cm}
    >{\raggedright\arraybackslash}p{5cm}
  }
    \toprule
    \textbf{Task} & \textbf{System Prompt} & \textbf{User Prompt} \\
    \midrule
    Generate Fortran unit tests & You're a proficient Fortran programmer. & Given Fortran code, write unit tests using funit.  
    Example: 
    \begin{minipage}{3cm}\vspace{2pt}\begin{verbatim}
FORTRAN CODE: [...]
FORTRAN TESTS: [...]\end{verbatim}\end{minipage}
\begin{minipage}{3cm}
    \vspace{3pt}
    Your turn: 
    \begin{verbatim}
FORTRAN CODE: {fortran_code}
FORTRAN TESTS:\end{verbatim} \end{minipage}\\
    \midrule
    Translate Fortran to Python & You're a programmer proficient in Fortran and Python. & Convert the following Fortran function to Python. \verb|```{python_code}```| \\
    \midrule
    Translate Fortran unit tests to Python & You're a programmer proficient in Fortran and Python. &  Convert the following unit tests from Fortran to Python using pytest. No need to import the module under test. \verb|```{unit_tests}```| \\
    \midrule
    Generate Python unit tests & You're a programmer proficient in Python and unit testing. You can write and execute Python code by enclosing it in triple backticks, e.g. ```code goes here``` & Generate 5 unit tests for the following Python function using pytest. No need to import the module under test. \verb|```{python_function}```| \\
\midrule
    Iteratively improve code & You're a programmer proficient in Fortran and Python. You can write and execute Python code by enclosing it in triple backticks, e.g. \verb|```code goes here```|.\par
When prompted to fix source code and unit tests, always return a response of the form:\par
SOURCE CODE: \verb|```<python source code>```|\par
UNIT TESTS: \verb|```<python unit tests>```|. Do not return any additional context.
 & Function being tested: \par \verb|{python_function}| \par
Here are some unit tests for the above code and the corresponding output. Unit tests: \verb|{python_unit_tests}| \par
Output from \verb|`pytest`|: \par
\verb|```| \{python\_test\_results\} \verb|```| \par

Modify the source code to pass the failing unit tests. Return a response of the following form: \par
SOURCE CODE: \verb|```<python source code>```| \par
UNIT TESTS: \verb|```<python unit tests>```| \\
    \bottomrule
  \end{tabular}
  \vspace{3pt}
  \caption{Prompts used for each task. Braces (\{\}) represent interpolated variables from the current translation task. Sections marked [...] were omitted for brevity. }
  \label{tab:prompts}
\end{table}

\subsection{Divide}

The goal of this step is to divide the large initial codebase into small problems, which can be solved by the `conquer` module in order. We have two important constraints here: the sub-problems have to be small enough that they fit within the LLM's context length, and they have to be individually testable. 

\begin{figure}[ht]
  \centering
  \includegraphics[width=0.75\textwidth]{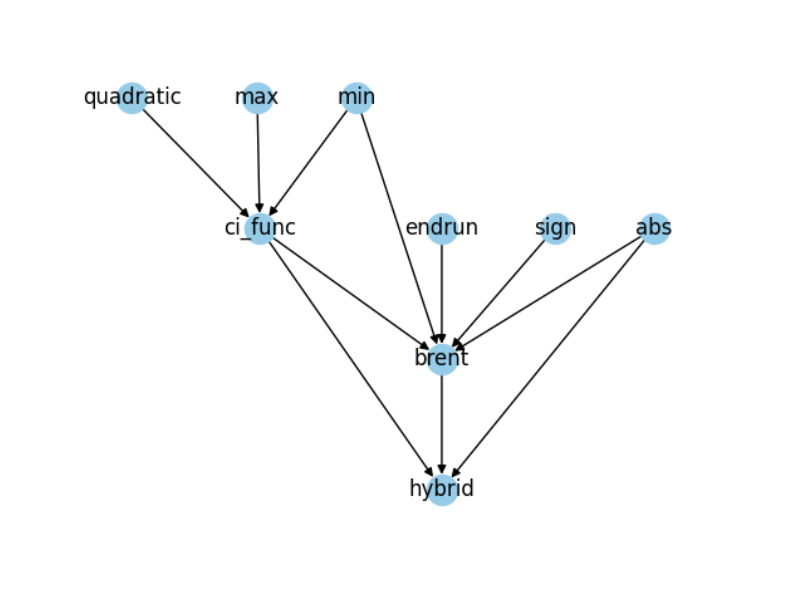}
  \caption{Dependency graph for the function `hybrid` from the leaf-level photosynthesis module. Each node corresponds to a function defined in this module, and each edge corresponds to a function call. }
  \label{fig:dep_graph}
\end{figure}

If we want the code chunks to be short enough for a model's context length, we should translate one function at a time (the whole module would be too large for ChatGPT). We also note that the functions must be translated in a particular order, because of their dependencies. For example, consider the functions represented by Figure \ref{fig:dep_graph}. In this case, we can't use the `hybrid` function until we've translated the other eight functions. 

Since we have here a directed acyclic graph (DAG) \citep{lipsky2022causal}, a topological sorting algorithm would yield a correct order for translating the functions one-by-one. This is exactly the approach we take for dividing the problem: generate a dependency graph of symbols, and then use topological sorting to determine an order of translation. For each unit of code, we then generate and run unit tests using GPT-4 using prompts from Table \ref{tab:prompts}. 

\subsubsection{Generating a dependency graph}

\begin{figure}[ht]
  \centering
  \includegraphics[width=1\textwidth]{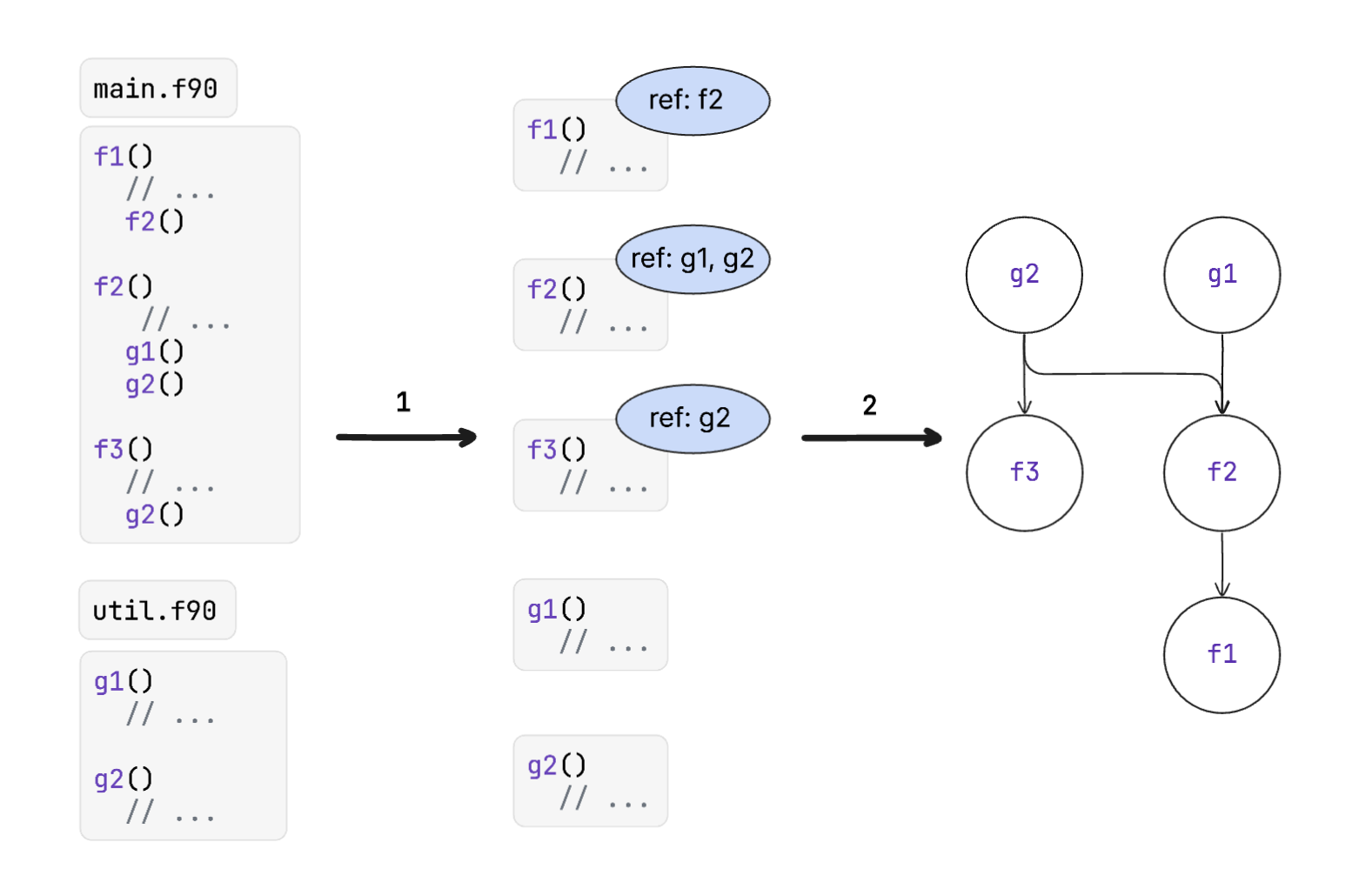}
  \caption{Visualization of code chunking process. In step 1, we chunk a codebase into individual units using a parsing tool and trace references. In step 2, we use those references to form a dependency graph.  }
  \label{fig:code_chunking}
\end{figure}

To create a dependency graph, we take a two-step approach. First, we divide the codebase into testable units (functions, types, or subroutines) and find the other units referenced by each unit. Then, we form a dependency graph based on these references. These steps are marked as step 1 and 2 in Figure \ref{fig:code_chunking}.

To chunk the code into units and find references, one method is using a parsing tool such as \verb|fparser| to return a syntax tree from the original source code. However, these tools only work on single files and can't unambiguously locate function definitions, requiring manual searches. So a parser is effective for chunking code but not for finding references.

To find references, there are at least two options. First, one could use the language server protocol (LSP), used in text editors like VSCode for features like "Go to Definition" \cite{lspgithub} \cite{fortls}. Second, one could prompt a language model, providing an index of all names in the project as context . This second approach has its limitations, because the index may exceed context length and generations may be unreliable. Given these downsides, LSP seems to be a good balanced approach for parsing dependencies, given its proven reliability and efficiency with codebases of any size. In this work, we created Python scripts using both LSP and fparser to compute dependency graphs. However, for even better results, using a compiler's data flow or control flow for ordered translation could be an exciting future direction. 

Our ultimate approach is documented in Figure \ref{fig:code_chunking}.

\subsubsection{Developing unit tests}

Once we use dependency graphs to create units, each of these units needs to be unit tested. In other words, we should be able to run a chunk of the Fortran code, along with the corresponding Python translation, and get the same outputs for a given set of inputs. To do this, we propose (as future work) implementing a logging tool, which would log the inputs and outputs for a Fortran function when it runs within the original code. A tool like \verb|kgen| would be good for this \cite{kgen2019}.  Then we could use these inputs and outputs to generate unit tests.  For now, we rely on GPT-4 to generate unit tests using its own knowledge, using the prompts from Table \ref{tab:prompts}.

\subsection{Conquer}

Assuming the divide step works well, we would want a consistent way to make GPT-4 write code that passes our unit tests. To do this, we have implemented an iterative approach to code generation, which takes in a chunk of Fortran code, along with corresponding unit tests, and generates corresponding Python code and Python unit tests. Then we make a Docker image that runs the tests automatically. If any tests cases fail, the test output gets passed back into the GPT-4 API, and it returns some revised code. This approach is depicted in Figure \ref{fig:workflow}.

\subsection{Combining divide and conquer}

In summary, we created a module for identifying the dependencies between symbols in Fortran source code, as well as a command-line interface for generating and iteratively updating a Python translation. 
To provide a proof of concept, here are some demonstrations of runtime and parameter estimation for the leaf-level photosynthesis module from the Community Land Model, leveraging the power of automatic differentiation for model parameter tuning.

\begin{table}[ht]
    \begin{adjustwidth}{-1in}{-1in}
    \begin{tabular}{|c|c|}
        \hline
        Fortran Source Code & Fortran Unit Tests \\
        \hline
        \begin{minipage}{0.6\textwidth}
            \tiny
            \begin{lstlisting}[language=Fortran]
elemental real(r8) function daylength(lat, decl)
    ! ... some comments omitted for conciseness
    use shr_infnan_mod, only : nan => shr_infnan_nan, &
                               assignment(=)
    use shr_const_mod , only : SHR_CONST_PI
    ! !ARGUMENTS:
    real(r8), intent(in) :: lat    
    real(r8), intent(in) :: decl 
    ! !LOCAL VARIABLES:
    real(r8) :: my_lat             
    real(r8) :: temp              
    ! number of seconds per radian of hour-angle
    real(r8), parameter :: secs_per_radian = 13750.9871_r8
    ! epsilon for defining latitudes "near" the pole
    real(r8), parameter :: lat_epsilon = 10._r8 * epsilon(1._r8)
    ! Define an offset pole as slightly less than pi/2 to avoid 
    ! problems with cos(lat) being negative
    real(r8), parameter :: pole = SHR_CONST_PI/2.0_r8
    real(r8), parameter :: offset_pole = pole - lat_epsilon

    ! lat must be less than pi/2 within a small tolerance
    if (abs(lat) >= (pole + lat_epsilon)) then
        daylength = nan
    ! decl must be strictly less than pi/2
    else if (abs(decl) >= pole) then
        daylength = nan
    ! normal case
    else    
        ! Ensure that latitude isn't too close to pole, to avoid 
        ! problems with cos(lat) being negative
        my_lat = min(offset_pole, max(-1._r8 * offset_pole, lat))
        temp = -(sin(my_lat)*sin(decl))/(cos(my_lat) * cos(decl))
        temp = min(1._r8,max(-1._r8,temp))
        daylength = 2.0_r8 * secs_per_radian * acos(temp) 
    end if
end function daylength
            \end{lstlisting}
        \end{minipage}
        & 
        \begin{minipage}{0.6\textwidth}
            \tiny
            \begin{lstlisting}[language=Fortran]
module test_daylength
  ! Tests of the daylength function in DaylengthMod
  use funit
  use shr_kind_mod , only : r8 => shr_kind_r8
  use shr_const_mod, only : SHR_CONST_PI
  use DaylengthMod , only : daylength
  implicit none
  save
  real(r8), parameter :: tol = 1.e-3_r8
contains
  @Test
  subroutine test_standard_points()
    ! Tests multiple points, not edge cases
    @assertEqual([26125.331_r8, 33030.159_r8], 
        daylength([-1.4_r8, -1.3_r8], 0.1_r8), 
        tolerance=tol)
  end subroutine test_standard_points
  @Test
  subroutine test_near_poles()
    ! Tests points near the north and south pole, which
    ! should result in full night and full day
    @assertEqual([0.0_r8, 86400.0_r8], 
        daylength([-1.5_r8, 1.5_r8], 0.1_r8), 
        tolerance=tol)
  end subroutine test_near_poles
  @Test
  subroutine test_edge_cases()
    ! Tests the edge cases, not the valid cases
    @assertEqual([1.e100_r8, -1.e100_r8], 
        daylength([1.5_r8, -1.5_r8], 0.1_r8), 
        tolerance=tol)
  end subroutine test_edge_cases
end module test_daylength
            \end{lstlisting}
        \end{minipage} \\
 \hline
        Python Source Code & Python Unit Tests \\
        \hline
        \begin{minipage}{0.6\textwidth}
            \tiny
            \begin{lstlisting}[language=Python]
import numpy as np


def daylength(lat, decl):
    """
    Calculate the length of the day (in hours) given the 
    latitude and the declination of the sun.  This is 
    the number of seconds between sunrise and sunset. 
    Returns NaN if input arguments are invalid.

    ... [more comments omitted for conciseness]
    """
    # Number of seconds per radian of hour-angle
    secs_per_radian = 13750.9871

    # Epsilon for defining latitudes "near" the pole
    lat_epsilon = 10.0 * np.finfo(float).eps

    pole = np.pi / 2
    offset_pole = pole - lat_epsilon

    # Lat must be less than pi/2 within a small tolerance
    # Decl must be strictly less than pi/2
    lat = np.where(abs(lat) >= pole + lat_epsilon, np.NAN, lat)
    decl = np.where(abs(decl) >= pole, np.NAN, decl)

    my_lat = np.clip(lat, -offset_pole, offset_pole)
    temp = -np.tan(my_lat) * np.tan(decl)
    temp = np.clip(temp, -1, 1)
    return 2.0 * secs_per_radian * np.arccos(temp)


class Bounds:
    def __init__(self, begg, endg):
        self.begg = begg
        self.endg = endg


def compute_max_daylength(bounds, lat, obliquity):
    """Compute max daylength for each grid cell"""
    max_daylength = []
    for g in range(bounds.begg, bounds.endg):
        max_decl = obliquity
        if lat[g] < 0.0:
            max_decl = -max_decl
        max_daylength.append(daylength(lat[g], max_decl))
    return max_daylength
            \end{lstlisting}
        \end{minipage} 
        & 
        \begin{minipage}{0.6\textwidth}
            \tiny
            \begin{lstlisting}[language=Python]
import numpy as np
import pytest
from daylength import daylength

# tolerance
tol = 1e-3

def test_standard_points():
    assert np.allclose(daylength(np.array([-1.4, -1.3]), 0.1), 
                       np.array([26125.331, 33030.159]), 
                       atol=tol)

def test_near_poles():
    assert np.allclose(daylength(np.array([-1.5, 1.5]), 0.1), 
                       np.array([0.0, 86400.0]), 
                       atol=tol)

def test_north_pole():
    assert abs(daylength(np.pi/2.0, 0.1) - 86400.0) < tol
    assert abs(daylength(np.pi/1.999, 0.1) - 86400.0) < tol

def test_south_pole():
    assert abs(daylength(-1.0 * np.pi/2.0, 0.1)) < tol
    assert abs(daylength(-1.0 * np.pi/1.999, 0.1)) < tol

def test_error_in_decl():
    assert np.isnan(daylength(-1.0, -3.0))

def test_error_in_lat_scalar():
    assert np.isnan(daylength(3.0, 0.1))

def test_error_in_lat_array():
    my_result = daylength(np.array([1.0, 3.0]), 0.1)
    assert np.isfinite(my_result[0])
    assert np.isnan(my_result[1])
            \end{lstlisting}
        \end{minipage}
        \\
        \hline
        
    \end{tabular}
    \end{adjustwidth}
    \caption{Sample outputs from a translation run on the day length function in the Community Land Model. Clockwise from top-left: original Fortran code, original Fortran unit tests, generated Python code, and generated Python unit tests. }
    \label{tab:outputs}
\end{table}

\end{document}